\documentclass[reprint,prl,showpacs,aps,amsmath,amssymb,nofootinbib,floatfix,floats,
superscriptaddress,twocolumn]{revtex4-1} 

\allowdisplaybreaks[2]
  
\usepackage[usenames, dvipsnames]{color}
\usepackage[colorlinks, pdfborder={0 0 0}, plainpages=false]{hyperref}
\usepackage{breakurl}
\usepackage{amsmath, amssymb, amsfonts}
\usepackage{xspace} % Sensible space treatment at end of simple macros
\usepackage{dcolumn}% Align table columns on decimal point
\usepackage{bm}% bold math
\usepackage{multirow} % Multiple rows in tables
\usepackage{graphics,psfrag}
\usepackage{graphicx,psfrag}
\usepackage{natbib}
\usepackage{longtable}
\usepackage{amsmath}
\usepackage{times}
\usepackage{mathptmx}

\include{macros}             % standard paper writing

\newcommand{\Caltech}{\affiliation{TAPIR, 
Walter Burke Institute for Theoretical Physics, California
Institute of Technology, Pasadena, CA 91125, USA}}
\newcommand{\JPL}{\affiliation{Jet Propulsion Laboratory,
    California Institute of Technology,
4800 Oak Grove Drive, Pasadena, CA 91106, USA}}
\newcommand{\Cornell}{\affiliation{Center for Radiophysics and Space
    Research, Cornell University, Ithaca, New York 14853, USA}}
\newcommand{\CITA}{\affiliation{Canadian Institute for Theoretical
    Astrophysics, 60 St.~George Street, University of Toronto,
    Toronto, ON M5S 3H8, Canada}} %
\newcommand{\CIFAR}{\affiliation{Canadian Institute for Advanced Research, 180 Dundas St.~West, Toronto, ON M5G 1Z8, Canada}} %

\newcommand{\Maryland}{\affiliation{Department of Physics,
    University of Maryland, College Park, MD 20742, USA}} %
\newcommand{\AEI}{\affiliation{Max Planck Institute for Gravitational Physics
(Albert Einstein Institute), Am M\"uhlenberg 1, Potsdam-Golm, 14476, Germany}} %
\newcommand{\Princeton}{\affiliation{Department of Physics, Princeton University, Jadwin Hall, Princeton, NJ 08544, USA}}

\begin{document}

\title{Numerical relativity reaching into post-Newtonian territory: a compact-object binary simulation spanning 350 gravitational-wave cycles}

%\author{B\'{e}la Szil\'{a}gyi} \Caltech % <bela@caltech.edu>
%\author{Jonathan Blackman} \Caltech
%\author{Tony Chu} \Princeton \CITA % <tonychu@princeton.edu>
%\author{Lawrence E.~Kidder} \Cornell % <kidder@astro.cornell.edu>
%\author{Harald P.~Pfeiffer} \CITA \CIFAR % <pfeiffer@cita.utoronto.ca>
%\author{Mark A.~Scheel} \Caltech % <scheel@tapir.caltech.edu>
%\author{Alessandra Buonanno} \AEI\Maryland %
%\author{Yi Pan} \Maryland %
%\author{Andrea Taracchini} \AEI%

\author{B\'{e}la Szil\'{a}gyi} \Caltech\JPL % <bela@caltech.edu>
\author{Jonathan Blackman} \Caltech
\author{Alessandra Buonanno} \AEI\Maryland %
\author{Andrea Taracchini} \AEI%
\author{Harald P.~Pfeiffer} \CITA \CIFAR % <pfeiffer@cita.utoronto.ca>
\author{Mark A.~Scheel} \Caltech % <scheel@tapir.caltech.edu>
\author{Tony Chu} \Princeton \CITA % <tonychu@princeton.edu>
\author{Lawrence E.~Kidder} \Cornell % <kidder@astro.cornell.edu>
\author{Yi Pan} \Maryland %

\date{\today}

\begin{abstract}
We present the \emph{first} numerical-relativity simulation of a compact-object binary whose gravitational waveform is long enough to cover the
  \emph{entire} frequency band of advanced gravitational-wave detectors, such as LIGO, Virgo and KAGRA, 
for mass ratio 7 and total mass as low as $45.5\,M_\odot$. We find that effective-one-body 
  models, either \emph{uncalibrated} or \emph{calibrated} against substantially shorter numerical-relativity waveforms at
  smaller mass ratios, reproduce our new waveform remarkably well, with a negligible loss in detection 
rate due to modeling error. 
In contrast, post-Newtonian inspiral waveforms and existing calibrated phenomenological inspiral-merger-ringdown waveforms display 
greater disagreement with our new simulation. The disagreement varies substantially depending on the specific 
post-Newtonian approximant used.
\end{abstract}

% 04.25.D- Numerical relativity
% 04.25.dg Numerical studies of black holes and black-hole binaries
% 04.25.Nx Post-Newtonian approximation; perturbation theory; related approximations
% 04.30.-w Gravitational waves (see also 04.80.Nn Gravitational wave detectors and experiments)
% 04.30.Db Wave generation and sources
% 02.70.Hm Spectral methods

\maketitle

{\it Introduction.} 
The upgraded ground-based interferometric gravitational-wave (GW)
detectors LIGO~\cite{aLIGO1,aLIGO2} and Virgo~\cite{AdV} 
will begin scientific
observations in mid 2015, and are expected to reach design sensitivity
by 2019~\cite{Aasi:2013wya}. Furthermore, a new Japanese detector,
KAGRA~\cite{Somiya:2011np}, is under construction.  Direct detection
of GWs by the end of this decade is therefore very likely. 
The most promising GW sources are compact-object
binaries, wherein each partner is either a stellar-mass black hole
(BH) or a neutron star (NS)~\cite{Abadie:2010cfa}.  The detection of
GWs from compact-object binaries, as well as 
 the determination of source properties from
detected GW signals, relies on the accurate knowledge of the
expected gravitational waveforms via matched-filtering~\cite{Finn1993} and
Markov chain Monte Carlo techniques.

The need for accurate waveforms has motivated
intense research.
Early
waveform models, based on the post-Newtonian 
(PN) formalism~\cite{Blanchet2006}, were limited to the early inspiral. 
The effective-one-body (EOB) formalism~\cite{Buonanno99,Buonanno00}
extended waveform models to the late inspiral, merger and ringdown.
Since 2005 research has greatly benefited from 
numerical-relativity (NR) simulations~\cite{Pretorius2005a,Campanelli2006a,Baker2006a}% 
\footnote{Besides its importance for GW astronomy, NR has also deepened the
understanding of general relativity in topics such as
binary BH recoil~\cite{Campanelli2007,Gonzalez2007} and gravitational
self force~\cite{Tiec:2013twa}.}. Current inspiral-merger-ringdown (IMR)   
waveform 
models~\cite{Ajith2009,Santamaria:2010yb,Damour:2012ky,Taracchini:2013rva} 
combine information from analytical-relativity (AR) calculations (best suited for the 
inspiral, when comparable-mass binaries  
have characteristic velocities smaller
than the speed of light) and direct NR simulations (the best means to 
explore the late 
inspiral and the merger).

\begin{table*}
  \centering
  \begin{tabular}{| c | c | c | c | c | c | c | c | c | c | c | }
    \hline
 \multicolumn{5}{|c|}{Initial Data} &   \multicolumn{4}{c|}{Inspiral} &   \multicolumn{2}{c|}{Remnant properties} \\
$D_{0}/M$ &  $10^3 M \Omega_{0}$ & $10^6\dot{a}_{0} M$ & $E_{\mathrm{ADM}}/M$ & $J_{\mathrm{ADM}}/M^2$ 
& $m_1/m_2$ & $10^5\epsilon$ & $T/M$ & $N$   & $M_{f}/M$  & $S_f/M_{f}^2$  \\ %&  $v(10^3km/s)$\\
    \hline
12.2 & 21.1541 & $-47.99$ & 0.996211 & 0.4510 & 6.99997(2) & $<6$ & $\;\;\;\,$4,100 & 20 & 0.98771(1)$\;\,$ & 0.32830(3)  \\
27   &  6.7930 & 0 & 0.998112 & 0.6123 & 7.00000(1) & 34 & 106,000 & 176 & 0.98762(14) & 0.32827(2) \\ %& $150(20)$  \\
    \hline
  \end{tabular}
  \caption{Properties of the two NR simulations: The first block lists initial separation $D_{0}$, orbital frequency $\Omega_{0}$, radial velocity $\dot a_{0}$, ADM energy $E_{\rm ADM}$ and angular momentum $J_{\rm ADM}$ in units of total mass $M=m_1+m_2$.  The middle block
lists
 mass ratio $m_1/m_2$, eccentricity $\epsilon$, 
time duration $T$ and number of orbits $N$ until merger. 
The final block lists remnant mass $M_f$ and spin $S_f$. 
}
\label{tab:properties}
\end{table*}

However, there is a gap between the portion of the binary evolution
that is described by analytical methods, and the portion that is
accessible by NR.  For example, waveforms computed 
at the currently-known PN order become unreliable
possibly hundreds of orbits before merger for unequal-mass
binaries~\cite{Damour:2010,MacDonald:2012mp}, and even
earlier when one of the objects is spinning~\cite{Nitz:2013mxa}~\footnote{Several PN waveforms (or approximants) with different PN-truncation error are available in the literature.  These PN approximants can differ from each other during 
the last hundreds of cycles before merger.} yet NR simulations have been able to cover only tens of
orbits~\cite{Buchman:2012dw,Mroue:2013PRL,Hinder:2013oqa} until now.
This gap has emerged as perhaps the most important source of uncertainty in
present IMR waveform models.
It is possible to construct IMR  
models by extending analytical waveforms across 
the gap~\cite{Ajith2009,Santamaria:2010yb,Damour:2012ky,Taracchini:2013rva},
 in some cases obtaining
IMR models that are faithful to longer numerical
waveforms when extrapolated beyond their limited range of
calibration~\cite{Pan:2013tva}. 
 However, so far these procedures have been tested using NR simulations with 
only 30 orbits, too few to close the gap.

The time duration $T$ of an inspiral waveform starting at initial GW
frequency $f_{\rm ini}$ scales as $T\!\propto\! \nu^{-1} f_{\rm
  ini}^{-8/3}$, where $\nu=m_1m_2/M^2$ is the symmetric mass ratio of
the binary with component masses $m_{1,2}$ and total mass $M=m_1+m_2$.
Therefore, reducing $f_{\rm ini}$ by a factor of $2$
increases $T$ sevenfold.  Halving the symmetric mass ratio
$\nu$ (e.g., from $m_1/m_2\!=\!2$ to $m_1/m_2\!=\!7$) doubles $T$.  
Increasing the simulation length $T$ is difficult: It becomes harder
to preserve phase coherency, the outer boundary of a simulation is in
causal contact for a larger fraction of the simulation, and existing
codes would require many months or even years of wall-clock time.
Therefore progress toward longer simulations
has been sluggish, with $T$ increasing by only about a factor of 2 to
3 during the last five years~\cite{Aylott:2009ya,Ajith:2012az,Hinder:2013oqa,Boyle2007,Mroue:2013PRL}. 
The duration $T$ needed to close the gap depends on the binary parameters 
and the detector bandwidth. Here we start addressing the issue of the gap 
by focusing on the nonspinning case and high mass ratio, $q=m_1/m_2=7$, for which the PN approximants 
can greatly differ~\cite{Damour:2000zb,MacDonald:2012mp}. We present a new NR simulation that extends
$T$ by a factor of $20$ and reduces the initial frequency $f_{\rm ini}$ by
a factor of $3$.  With its comparatively high mass ratio,  
the new simulation probes an astrophysically relevant parameter regime 
for BH-BH and NS-BH binaries and for certain total masses covers the entire 
frequency band of advanced LIGO (aLIGO) and Virgo.  We describe challenges
involved in carrying out this new simulation, 
most notably an instability
that causes the center of mass (CoM) of the binary to move, and we suggest
improvements for future long simulations.  
We then compare the new simulation with
existing analytical waveform models to assess 
the impact of waveform model errors on the detection rate of 
advanced detectors. 

\begin{figure}
\begin{center}
\includegraphics[width=0.9\columnwidth,trim=13 29 1 1,clip=true]{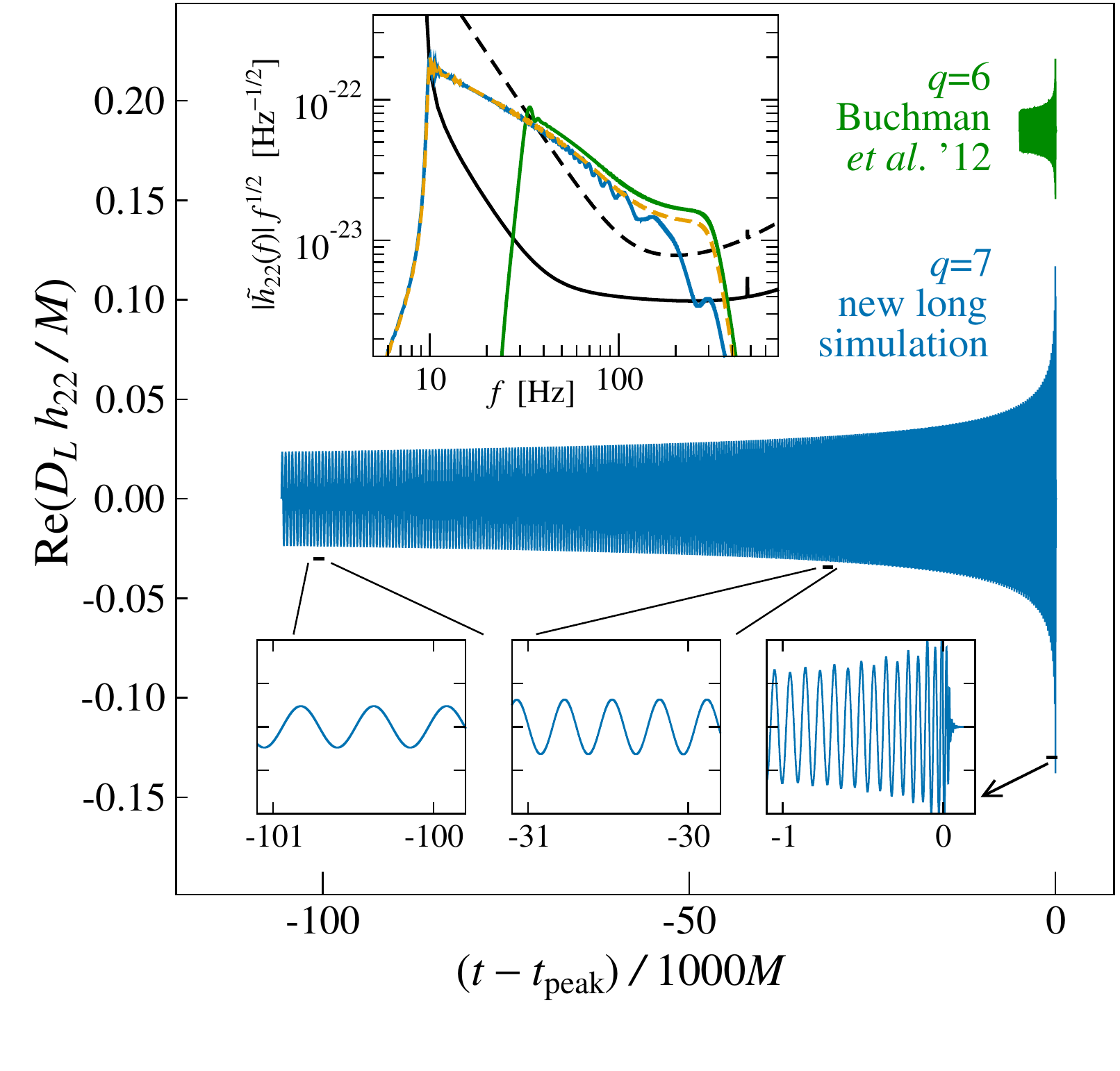}
\caption{\label{fig:waveform} 
Overview of the new very long simulation.  The main panel 
shows the $(2,2)$ spherical-harmonic mode of the GW strain, 
with enlargements in the lower insets.
The top inset shows the Fourier spectra of the new waveform in blue
and the NR-NR hybrid waveform (used for comparisons with analytical models) in
yellow, overlaid with noise power spectral densities of aLIGO at the
early (dashed black) and design (solid black)
sensitivity~\cite{Aasi:2013wya}.  The waveforms in the inset are
scaled to total mass $M=45.5\,M_\odot$ and luminosity distance
$D_L\approx1.06$\,Gpc.
For comparison, an older $q=6$ waveform~\cite{Buchman:2012dw} 
of representative length is shown in the main 
panel (offset vertically for clarity) and in the
power-spectrum inset. 
}
\end{center}
\end{figure}

{\it Numerical-relativity waveforms.} We report on
two new simulations of a nonspinning BH binary with mass ratio $q=m_1/m_2=7$.  
The \emph{short} simulation is of
typical length: 20 orbits, $T=4,100\,M$.  
The \emph{long} simulation, the main focus of this paper, is about 25 times longer. 
Both simulations are computed using the Spectral Einstein Code 
({\tt SpEC})~\cite{SpECwebsite}. 
The short simulation uses established computational 
techniques~\cite{Mroue:2013PRL}. 
The speed-up needed for the 
long simulation is the result of a series of
code changes including task-based 
adaptive parallel load-balancing,
live timing-based selection of the most efficient algorithm
(when multiple implementations of the same function are available),
a modified memory layout to allow more efficient calls to 
low-level numerical packages 
and
a more efficient implementation of the Generalized Harmonic 
evolution equations. 
Figure~\ref{fig:waveform} shows the new long waveform and 
Table~\ref{tab:properties} presents additional details about both simulations. 
Geometrized units $G=c=1$ are used in Table~\ref{tab:properties} and throughout 
this paper. The top inset of Fig.~\ref{fig:waveform} shows 
the spectra of the $(2,2)$ spherical harmonic waveform modes. 
The long simulation covers the entire design-aLIGO frequency 
range for nonspinning BH-BH binaries with 
$M\gtrsim 45\,M_\odot$, and covers the early-aLIGO frequency range for 
$M\gtrsim 11\,M_\odot$, including nonspinning NS-BH binaries\footnote{For mass ratio 7, in absence of spin, we expect no observable 
differences in the merger signal between a BH-BH and a NS-BH binary~\cite{Foucart:2013psa}.}. In contrast, the $q=6$ simulation
 plotted in green, which is
representative of past simulations, starts at
  3 times higher frequency, and covers a much smaller portion of
    the aLIGO bandwidth for a given $M$.
Thus, we present here
 the first gravitational waveform covering the entire design-aLIGO 
frequency band for a nonspinning, 
compact-object binary at mass ratio $q=7$ with 
a total mass as low as $M=45.5\,M_\odot$.

The short simulation is run at three different
numerical resolutions, and the long one at four resolutions. 
The long simulation employs dynamical spectral adaptive mesh 
refinement~\cite{Szilagyi:2014fna}, so 
measured quantities (like BH masses or waveforms) do not always 
converge in a regular, predictable manner with increasing
resolution, as is the case when each resolution 
is defined by a fixed number of grid points.  Furthermore, failure to
resolve initial transients caused by imperfect initial data also
complicates convergence (see discussion in Sec.~IIIB of 
Ref.~\cite{Scheel2014}). Nevertheless,
we find that differences in measured quantities like waveforms, masses,
and spins decrease rapidly with resolution,
and Table~\ref{tab:properties} displays a conservative
error estimate obtained
by taking the difference between the two highest resolutions.
After the initial transients have
decayed, we measure the mass ratio to be equal to 7 to within five
significant digits, and the dimensionless spins to be $\lesssim 10^{-6}$.  
The remnant mass and spin agree to within four significant digits
between the short and long simulations.

\begin{figure}
\begin{center}
\includegraphics[width=0.98\columnwidth,trim=2 2 0 0,clip=true]{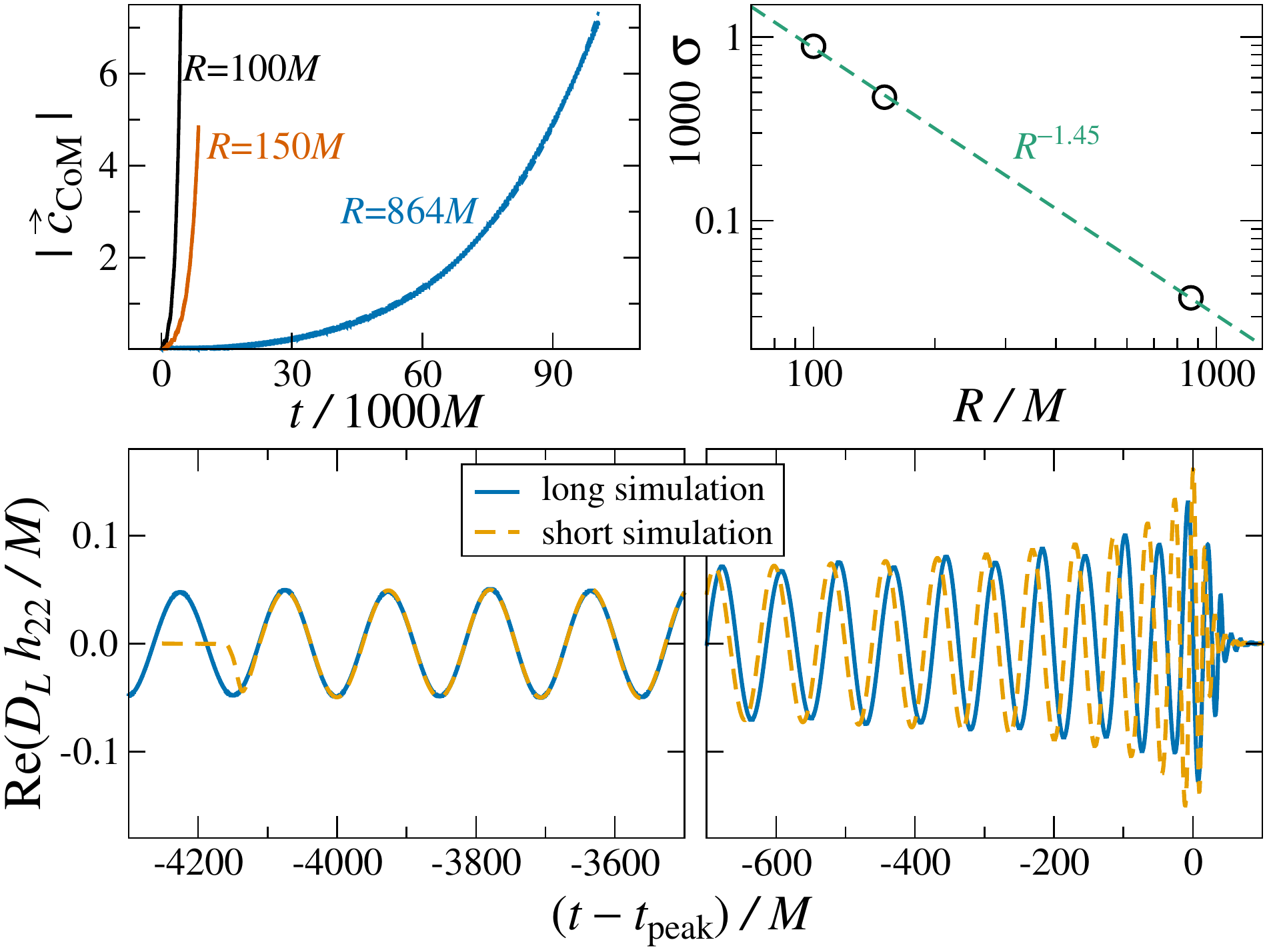}
\caption{\label{fig:drift} 
    {\bf Top
        left:} displacement of the CoM from the origin,
      for three long simulations with different outer boundary radii
      $R$.  In each case, $|\vec c_{\rm CoM}|$ increases
      exponentially.  {\bf Top right:} growth rate $\sigma$ of
      $|\vec{c}_{\rm CoM}|$ as a function of $R$ with a power-law fit.  
      {\bf Lower panels:}
      GW (2,2) mode of the short and long simulations.  
      The long simulation still agrees very well with the short
        one at early times, but fails to produce an accurate merger
        waveform.
}
\end{center}
\end{figure}

However, the long 
simulation encounters an unexpected problem.
After a few $10,000\,M$ of evolution, the coordinate CoM begins
to drift away from the origin.  Define the CoM  as
$ \vec{c}_{\rm CoM} = \vec c_1 {m_1}/{M}+ \vec c_2 {m_2}/{M}$,
where $\vec c_{1,2}$ are the coordinate centers of the apparent
horizons of the two BHs.  The top left panel of
Fig.~\ref{fig:drift} shows that $\lvert \vec c_{\rm CoM}\rvert$ 
increases {\em exponentially} with time, contradicting all known physical
expectations.  This drift is primarily a coordinate effect 
that only marginally affects most measurable quantities. For example,
the linear momentum radiated to infinity, as computed from the
waveform obtained by Cauchy-Characteristic extraction~\cite{Taylor:2013zia}, is
consistent with PN theory, and is too small to account for the
motion of the CoM. To explore the drift in more detail, we repeat the
long simulation with different values of $R$,
the coordinate radius of the artificial
outer boundary where we impose an outgoing-wave boundary condition.
The top left panel of Fig.~\ref{fig:drift} shows
the CoM drift for several values of $R$.  
We find that the exponential growth rate 
$\sigma$ behaves approximately like $\sigma \propto R^{-1.45}$,
as shown in the upper right panel of Fig.~\ref{fig:drift}.  For
our standard choice of $R=864\,M$, 
$1/\sigma = 26,000\,M$; this large timescale
explains why the drift was not noticed in earlier,
shorter simulations.  

We conjecture that the drift is caused by a coupling with the
  outer boundary.  Such a coupling might arise through enhanced
  reflections of the outgoing GW at the outer
  boundary.  Our outgoing-wave boundary
  conditions~\cite{Buchman2006} have the smallest reflection 
  coefficient for spherical
  harmonic modes with small $\ell$, and a 
  reflection coefficient of order unity when 
  $kR/\ell\gtrsim 1$~\cite{Buchman2006}, where $k$ is the radial wavevector.  
  With increasing $\lvert \vec c_{\rm
    CoM}\rvert$ the emitted GW will have increasing
  high-$\ell$ content when decomposed on the outer boundary.

From the bottom panel in Fig.~\ref{fig:drift}, we see that the effects
of the drift on the long waveform is confined to the last
$\sim10$ orbits before merger. In the Fourier domain (see
Fig.~\ref{fig:waveform}) one can clearly see the unusual behavior of
the long waveform at high frequencies. Since these orbits are covered by the short waveform, we
hybridize the short waveform with the 
long one, thus replacing
the final portion of the former. We adopt the hybridization
method of Ref.~\cite{MacDonald:2011ne}. We construct 9 NR-NR hybrids
by combining 3 
versions of the long simulation and three versions of
the short simulation.  Each version may differ by the numerical
resolution (which we label by ``Lev${\cal N}$",
 with larger integers ${\cal N}$
indicating finer resolution), or by the degree of
the polynomial used to extrapolate the waveform to 
infinity~\cite{Boyle-Mroue:2008,Taylor:2013zia} 
(which we label by ``N${\cal M}$", where ${\cal M}$ is the polynomial
degree.)
In particular, we use (Lev3,
N3), (Lev3, N2), and (Lev2, N3) for the long simulation, while
we use (Lev5, N3), (Lev5, N2), and (Lev4, N3) for the short
simulation.
The fiducial NR-NR hybrid is built from the long
(Lev3, N3) and the short 
(Lev5, N3) simulations; this
pair of waveforms is
blended over the interval $t-t_{\rm peak}\in [-3252,
-2252]M$. In the top panel of Fig.~\ref{fig:waveform}, we show in
yellow the spectrum of the fiducial NR-NR hybrid. This spectrum
behaves as expected close to merger, and is devoid of oscillations,
just like the spectrum of the $q=6$ simulation.
Since we cannot estimate the impact of the coordinate drift on the
phase error of the long waveform, we cannot make statements
about the phase disagreement between the long 
waveform and
analytical waveform models.  Nevertheless, we can compare the
analytical models to the hybrid NR-NR waveform, and investigate how
the results change when we vary the blending window where the
hybridization is done.

\begin{figure*}
%\vspace{-1cm}
\begin{center}
\includegraphics[width=0.9\columnwidth]{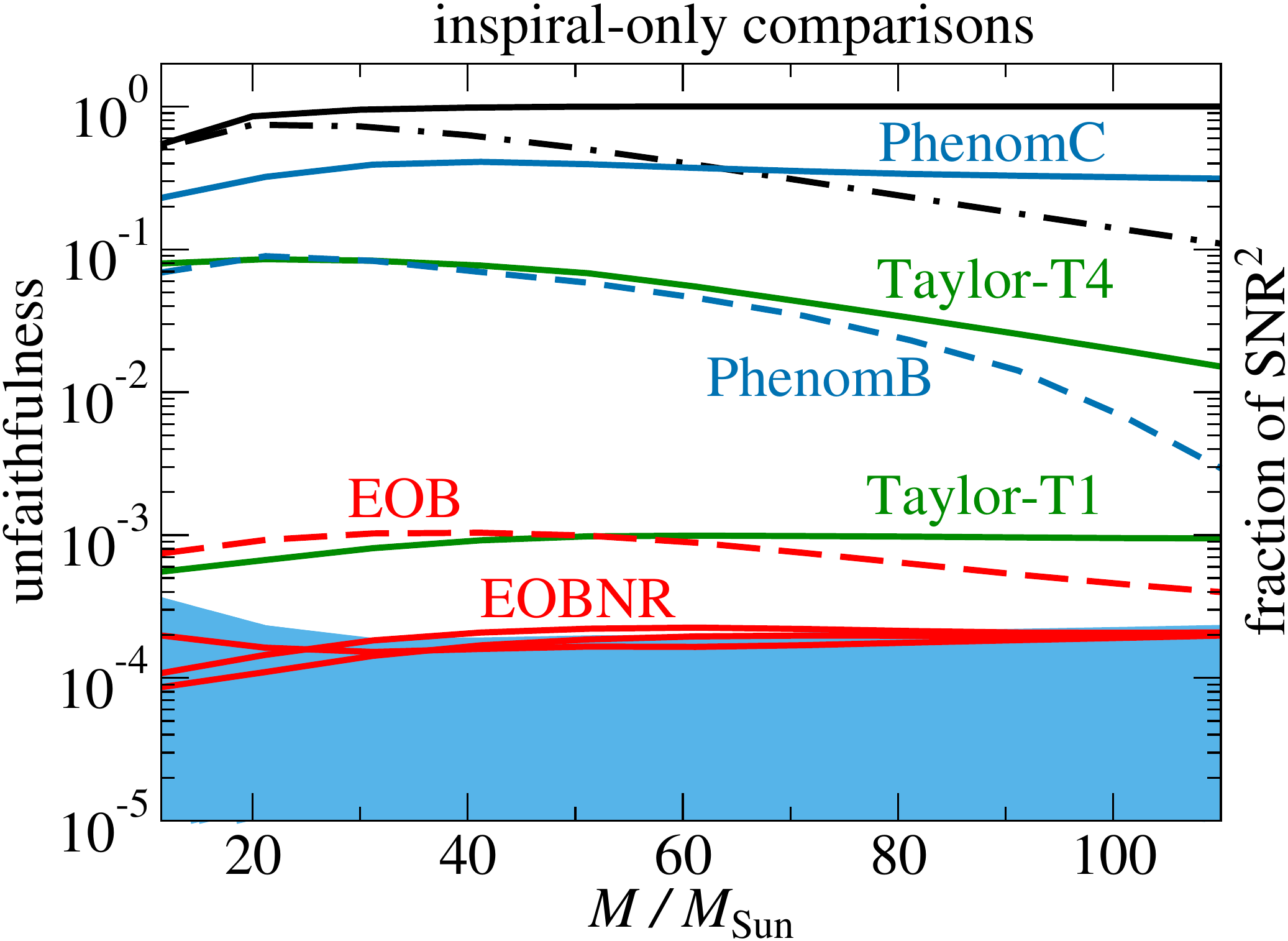}
%\vskip1ex
\hspace{0.5cm}
\includegraphics[width=0.9\columnwidth]{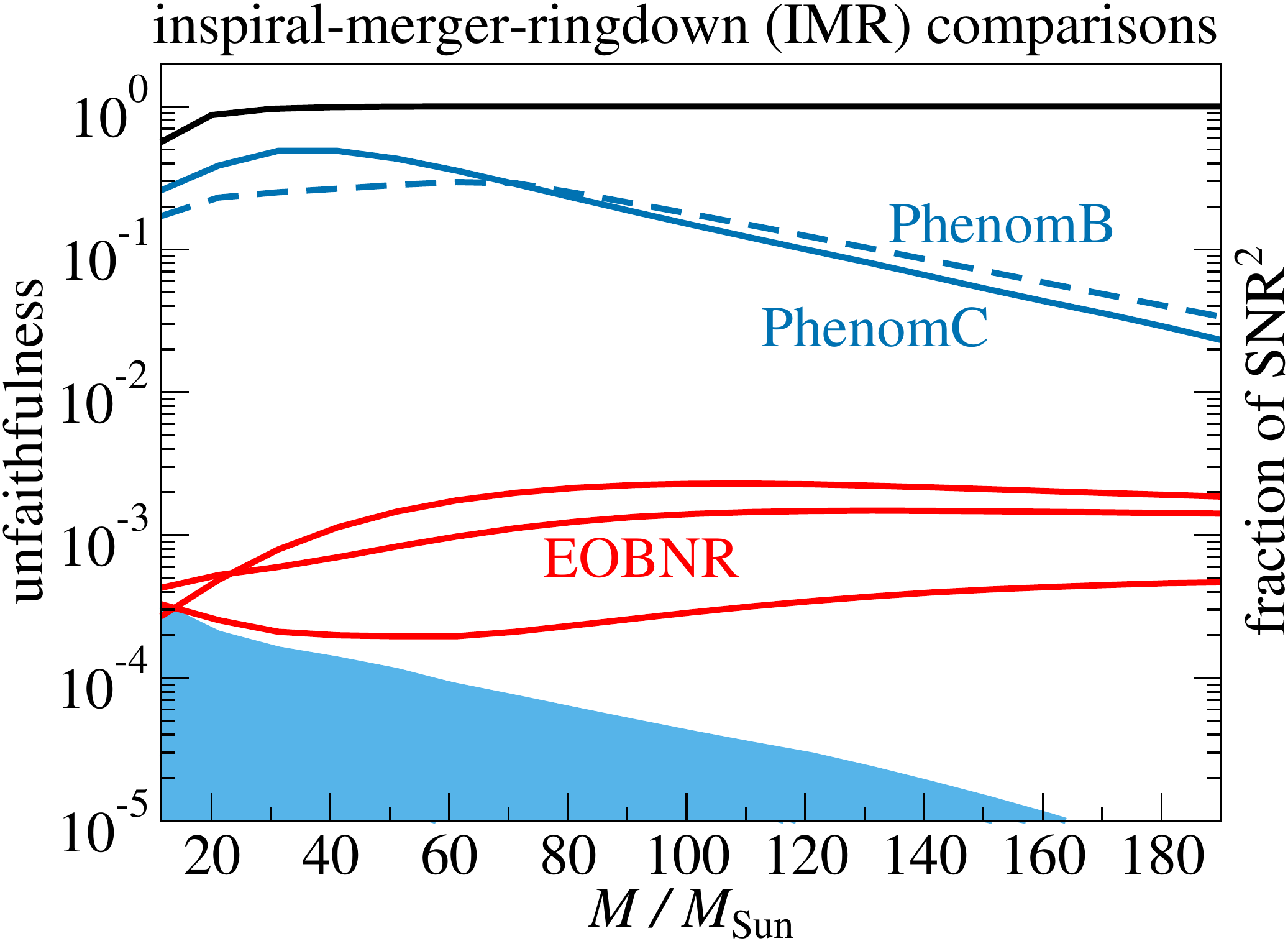}
  \caption{Unfaithfulness of the hybrid NR-NR waveform against several
    analytical waveform models. \textbf{Left panel:} inspiral-only comparisons.  \textbf{Right panel:} IMR comparisons.  Also shown in the left panel are SNR$_{\rm insp}^2$/SNR$_{\rm full-insp}^2$  (solid black line) and 
SNR$_{\rm insp}^2$/SNR$_{\rm full-IMR}^2$ (dot-dashed black line), and in the right panel SNR$_{\rm IMR}^2$/SNR$_{\rm full-IMR}^2$ 
(solid black line).  The blue area indicates the NR error.}
\label{fig:unfaith}
\end{center}
\end{figure*}

{\it Comparison to analytical-relativity waveforms.} 
Figure~\ref{fig:unfaith} summarizes our comparisons between various 
analytical waveforms $h_{22}^{\rm AR}$ with the hybrid NR-NR waveform
$h_{22}^{\rm NR}$.  Shown is the unfaithfulness $\bar{\mathcal F}$, defined
as $\bar{\mathcal F}\equiv 1 - \max_{t_{0},\phi_{0}} {\langle h_{22}^{\rm
    AR},h_{22}^{\rm NR}\rangle}/||h_{22}^{\rm AR}||/||h_{22}^{\rm
  NR}||$. Here
 $t_{0}$ and $\phi_{0}$ are the initial time and phase,
$||h||\equiv\sqrt{\langle h,h\rangle}$, and $\langle h_1,h_2\rangle
\equiv 4\, {\rm Re} \int_{f_{\rm min}}^{f_{\rm max}}
\tilde{h}_1(f)\tilde{h}^*_2(f)/S_n(f)\,{\rm d}f$, where $S_n(f)$ is
the zero-detuned, high-power noise power spectral density of
aLIGO~\cite{Shoemaker2009}. 
We consider the following analytical waveform models from the LIGO
Algorithm Library (LAL): the inspiral-only PN Taylor
approximants~\cite{Buonanno:2009} in the
time domain (Taylor-T1, T2, T4)
and in the frequency domain (Taylor-F2), an inspiral EOB
model (obtained from Ref.~\cite{Taracchini:2013rva} by dropping any NR information, thus uncalibrated), the IMR
EOBNR models that were obtained by calibrating 
the EOB model to NR simulations~\cite{Pan:2011gk,Taracchini:2012, Taracchini:2013rva} (denoted in LAL as EOBNRv2, SEOBNRv1 and SEOBNRv2), and the
IMR phenomenological models that were built combining PN and 
NR results~\cite{Ajith2009, Santamaria:2010yb} (denoted in LAL as PhenomB and PhenomC).  All the time-domain IMR waveforms are tapered using a Planck windowing
function~\cite{McKechan:2010kp}, both at the beginning and at the end. 
We generate the model waveforms starting from an initial GW frequency of $M\omega_{22}=0.01317$. 
For inspiral-only models, we set $f_{\rm max} = 0.01176/M$, the minimum available final GW frequency among the time-domain
Taylor models, a value close to the
innermost-circular-orbit value in Schwarzschild spacetime ($\approx 0.01083/M$), whereas for the IMR comparisons $f_{\rm max}=\infty$. 
Quite interestingly, the inspiral-only comparisons give
  similar results when employing directly the long simulation
  instead of the NR-NR hybrid.

The blue shaded area in Fig.~\ref{fig:unfaith} represents
the uncertainty in the NR waveforms, estimated by computing $\bar{\mathcal F}$ 
between the fiducial hybrid NR-NR waveform and the other 8 NR-NR hybrids. 
Because
the inspiral-only and IMR curves are calculated using different
portions of the hybrid NR-NR waveform, 
the same model may have different values in the two panels
for the same total mass. We vary the
prescriptions used for the hybridization (namely, position and width
of the blending window), and we find changes $\mathcal{O}(10^{-4})$
in the unfaithfulness curves for low total masses. Thus, we consider 
our results robust. If general relativity correctly describes the GW signals found in nature, 
then the unfaithfulness $\bar{\mathcal F}$ plotted in Fig.~\ref{fig:unfaith} yields
a bound on the loss in detection rate due to modeling error. 
For sources uniformly distributed in space, the relative loss in 
detection rate is $\sim 3(d_{\rm MM}  + d_{\rm E})$ (see 
Sec.~VB in Ref.~\cite{Buonanno:2009}) 
where $d_{\rm MM}$ is the minimal match of the template bank and 
$d_{\rm E}$=$1- \max_{\vec{\lambda}} {\langle h_{22}^{\rm
    AR},h_{22}^{\rm NR}\rangle}/||h_{22}^{\rm AR}||/||h_{22}^{\rm
  NR}||$ is the ineffectualness. Here $\vec{\lambda}$ describes {\it all} the binary parameters, 
not just $\phi_0$ and $t_0$, and therefore $d_{\rm E}\le \bar{\mathcal F}$. Typically, $d_{\rm MM} = 3\%$ in LIGO searches. 
Thus, to achieve $\lesssim 10\%$ loss in detection rate, 
it suffices that $\bar{\mathcal F}\lesssim 1\%$~\cite{Buonanno:2009}.

Quite remarkably, we find that the unfaithfulness of the uncalibrated  
{\it inspiral} EOB waveform is $<0.1\%$, with a negligible loss in
detection rate due to modeling error. The agreement is of course better for the
inspiral EOBNR waveforms (i.e., EOBNRv2, SEOBNRv1, SEOBNRv2) 
($\bar{\mathcal F} < 0.02\%$, left panel), and $\bar{\mathcal F}<0.2\%$
for the IMR EOBNR waveforms (right panel). The closeness of all
inspiral EOBNR waveforms strongly suggests that the different
calibrations and variations in the dynamics and energy fluxes of 
those EOBNR models~\cite{Pan:2011gk,Taracchini:2012,Taracchini:2013rva} do not
impact the low-frequency part of the waveforms, but affect (in a minor
way) only the last stages of the inspiral and the merger. The unfaithfulness
of the time- and frequency-domain inspiral-only PN Taylor approximants varies
between $0.1\%$ and $10\%$ depending on the binary's total mass and
the PN approximant used~\footnote{The large unfaithfulness 
of some of the PN Taylor approximants is due to differences in the evolution of 
the frequency and its first time derivative during the late inspiral 
phase.}. In particular, Taylor-T4, which has the best
agreement with NR in the equal-mass case~\cite{Boyle2007}, has the
largest disagreement with the new long $q=7$ NR waveform.
(The approximants Taylor-T2 and Taylor-F2 are not displayed, but lie
between Taylor-T1 and Taylor-T4).  
The PhenomB and C models were fitted 
to hybrids built with Taylor PN approximants and less accurate, short NR waveforms, 
which may in part explain the large disagreement we find.

The new long NR waveform covers the entire design-aLIGO
  frequency band only for total mass $M\ge 45.5M_\odot$; for smaller
  $M$, the unfaithfulness calculations in Fig.~\ref{fig:unfaith} neglect the
  lowest frequency portion of the waveform visible to aLIGO, down to $\sim 10$\,Hz. 
  To understand the significance of the missing GW cycles in the
  low-frequency portion of the bandwidth, we compute the signal-to-noise ratio (SNR) 
  accumulated within the frequency range of our comparisons
  ($\textrm{SNR}_{\rm insp}$ and $\textrm{SNR}_{\rm IMR}$ for the left
  and right panel of Fig.~\ref{fig:unfaith}, respectively), and
  compare it with the SNR accumulated over the \emph{entire}
  inspiral ($\textrm{SNR}_{\rm full-insp}$) and the \emph{entire} IMR ($\textrm{SNR}_{\rm full-IMR}$). 
To cover the entire design-aLIGO bandwidth we use the calibrated EOB model of Ref.~\cite{Taracchini:2013rva}.
  Suitable (squared) ratios of these quantities, which represent the fraction of total SNR that is 
 accessible to our comparisons, are plotted in Fig.~\ref{fig:unfaith}. These ratios are $<1$ whenever GW cycles are missing in the
range $10\,\textrm{Hz} \leq f \leq f_{\rm ini}^{\rm NR}$ or, in the case of
$\textrm{SNR}_{\rm insp}^2/\textrm{SNR}_{\rm full-IMR}^2$, also when
the merger-ringdown signal is in band. We find that, even for total
masses $<45.5\,M_\odot$, the unfaithfulness can still be a meaningful 
assessment of the quality of the analytical models, since a large fraction of SNR is accumulated. 
Because the merger-ringdown portion of the waveforms becomes
increasingly important at higher masses, the inspiral-only comparisons
at high mass cover only a small fraction of the entire SNR, as illustrated
by the steep decline of $\textrm{SNR}^2_{\rm insp}/\textrm{SNR}^2_{\rm full-IMR}$ in the left panel of Fig.~\ref{fig:unfaith}.

{\it Conclusions.}
To detect and extract unique, astrophysical information from 
coalescing compact-object binaries, GW instruments employ model waveforms 
built by combining analytical and numerical-relativity predictions~\cite{Ajith2009,Santamaria:2010yb,Damour:2012ky,
Taracchini:2013rva}.  Currently, the main uncertainty of those waveform models is caused by the gap between the 
regimes of applicability of those methods. This uncertainty can be addressed and, eventually, 
solved by running much longer NR simulations. In this work we start to tackle this 
issue by producing a BH-BH simulation $20$ times longer than previous simulations.
Because of an unexpected drift of 
the CoM during the last $40$ GW cycles, we construct the full NR inspiral, merger 
and ringdown waveform by hybridizing the long NR waveform with a new, short NR simulation. 
The hybrid NR-NR waveform covers the entire band of advanced GW detectors for 
total mass $\ge  45.5\,M_\odot$. Comparing to analytical waveform models, we find 
strong evidence that --- at least for nonspinning binaries --- the EOB formalism accurately
describes the inspiral dynamics in the so-far unexplored regime of 20
to 176 orbits before merger, and combined with previous work \cite{Pan:2013tva}  
provides accurate waveforms beyond the limited range of calibration. Quite remarkably, 
the excellent agreement of EOBNR waveforms holds also for uncalibrated inspiral EOB 
waveforms. PN-approximants have larger errors and more importantly  the errors vary substantially 
depending on the specific PN approximant used.

{\it Acknowledgments.} We thank Alejandro Boh\'e for useful discussions. 
A.B. acknowledges partial support from NSF Grant No. PHY-1208881 and NASA Grant NNX12AN10G. 
T.C. and H.P. gratefully acknowledge support from NSERC of
Canada, the Canada Chairs Program, and the Canadian Institute for
Advanced Research.  L.K. gratefully acknowledge
support from the Sherman Fairchild Foundation, and from NSF grants
PHY-1306125 and AST-1333129 at Cornell.
J.B. gratefully acknowledges support from NSERC of Canada.
M.S., B.Sz., and J.B. acknowledge support from the 
Sherman Fairchild
Foundation and from NSF grants PHY-1440083 and AST-1333520 
at Caltech. Simulations used in this work were
computed with the \texttt{SpEC} code~\cite{SpECwebsite}.  Computations
were performed on the Zwicky cluster at Caltech, which is supported by
the Sherman Fairchild Foundation and by NSF award PHY-0960291; on the
NSF XSEDE network under grant TG-PHY990007N; 
on the Orca cluster supported by Cal State Fullerton; 
and on the GPC supercomputer at the SciNet HPC Consortium~\cite{scinet}. SciNet is
funded by: the Canada Foundation for Innovation under the auspices of
Compute Canada; the Government of Ontario; Ontario Research
Fund--Research Excellence; and the University of Toronto.

%merlin.mbs apsrev4-1.bst 2010-07-25 4.21a (PWD, AO, DPC) hacked
%Control: key (0)
%Control: author (8) initials jnrlst
%Control: editor formatted (1) identically to author
%Control: production of article title (-1) disabled
%Control: page (0) single
%Control: year (1) truncated
%Control: production of eprint (0) enabled
%

%\bibliography{References/References}
%\begin{thebibliography}{99}
%\end{thebibliography}

\end{document}